\documentclass[useAMS,usenatbib]{mn2e}
\usepackage{times}
\usepackage{amsmath}%
\usepackage[pass]{geometry}
\usepackage{amsfonts}%
\usepackage{amssymb}%
\usepackage{graphicx}
\usepackage[english]{babel}
\usepackage[latin1]{inputenc}
\usepackage{multicol}
\usepackage{natbib}

\title{Parasitic interference in nulling interferometry}
\author[A. Matter et al.]{A.~Matter$^{1}$\thanks{E-mail: alexis.matter@obs.ujf-grenoble.fr (AM)}\thanks{Present address: Institut de plan\'etologie et d'astrophysique de Grenoble, 414 rue de la Piscine, Domaine universitaire, F-38400 Saint Martin d'H\`eres, France}, D.~Defr\`ere$^{1,5}$\footnotemark[1]\thanks{Present address: University of Arizona, 933 N. Cherry Ave., Tucson, AZ 85721, USA}, W.C.~Danchi$^{2}$\footnotemark[1], B.~Lopez$^{3}$\footnotemark[1], and O.~Absil$^{4}$\footnotemark[1]\\
$^{1}$Max Planck Institut für Radioastronomie, auf dem Hügel, 69, Bonn, 53121, Germany\\
$^{2}$NASA/GSFC, Greenbelt, MD 20771, USA\\
$^{3}$Laboratoire Lagrange, CNRS UMR 7293, UNS - Observatoire de la Côte d'Azur BP 4229, F-06304 Nice Cedex 4, France\\
$^{4}$D\'epartement d'Astrophysique, G\'eophysique et Oc\'eanographie, Universit\'e de Li\`ege, 17 All\'ee du Six Ao\^ut, B-4000 Li\`ege, Belgium\\
$^{5}$Steward Observatory, Department of Astronomy, University of Arizona, 933 N. Cherry Ave, Tucson, AZ 85721, USA}
\begin{document}
\bibliographystyle{mn2e}

\date{In original form 2013 March 05}

\pagerange{\pageref{firstpage}--\pageref{lastpage}} \pubyear{2013}

\maketitle

\label{firstpage}

\begin{abstract}
Nulling interferometry aims to detect faint objects close to bright stars. Its principle is to produce a destructive interference along the line-of-sight so that the stellar flux is rejected, while the flux of the off-axis source can be transmitted. In practice, various instrumental perturbations can degrade the nulling performance. Any imperfection in phase, amplitude, or polarization produces a spurious flux that leaks to the interferometer output and corrupts the transmitted off-axis flux. One of these instrumental pertubations is the crosstalk phenomenon, which occurs because of multiple parasitic reflections inside transmitting optics, and/or diffraction effects related to beam propagation along finite size optics. It can include a crosstalk of a beam with itself, and a mutual crosstalk between different beams. This can create a parasitic interference pattern, which degrades the intrinsic transmission map - or intensity response - of the interferometer. In this context, we describe how this instrumental effect impairs the performance of a Bracewell interferometer. A simple formalism is developed to derive the corresponding modified intensity response of the interferometer, as a function of the two parameters of interest: the crosstalk level (or contamination rate) and the phase shift between the primary and secondary - parasitic - beams. We then apply our mathematical approach to a few scientific cases, both analytically and using the \emph{GENIEsim} simulation software, adapted to handle coherent crosstalk. Our results show that a coherent crosstalk level of about 1\% implies a 20\% drop of the SNR at most. Careful attention should thus be paid to reduce the crosstalk level inside an interferometric instrument and ensure an instrumental stability that provides the necessary sensitivity through calibration procedures.  
\end{abstract}


\section{Introduction}
\citet{1978Natur.274..780B} developed the concept of nulling interferometry that aims at detecting faint off-axis sources - e.g., planets, exozodiacal discs - orbiting distant stars. Its principle is to enhance the companion over star flux ratio by producing a destructive interference on the line-of-sight so that the stellar flux is rejected. By an appropriate choice of baseline length and orientation, the flux of the off-axis source can be transmitted and thus detected more easily owing to the reduced photon noise from the on-axis star. \\ 
In practice, the rejection rate of the stellar flux is not perfect because of the finite size of the stellar photosphere, which causes the so-called ``geometric leakage''. Moreover, this rejection rate can be degraded by various instrumental effects such as imperfect co-phasing of the light beams, wavefront errors, intensity mismatches, and polarisation errors \citep{ollivier1999}. This contribution, called instrumental leakage, adds to the geometric leakage at the destructive output of the interferometer, and is generally considered as the main source of noise in nulling interferometry.\\ 
The impact of instrumental leakage on the detection of faint companions has been considered both experimentally \citep{2006ApOpt..45..984C,2010A&A...520A..96M}, and analytically \citep{2004ApOpt..43.6100L}. However, none of these studies addressed the coherent crosstalk that may occur between beams inside an interferometric instrument. 
In fact, for most of the current classical and nulling interferometers, the beams coming from each telescope are carried through tunnels up to a combining device. The beams are often reduced in size for practical reasons, and the current instruments are characterized by optical modules performing various functions such as spatial filtering, spectral band separation, and spectral resolution. However, the transport of these beams through multiple optical modules can be problematic. Because of diffraction effects associated with beam propagation along finite size optics, a coherent crosstalk may occur between beams. Occuring during the beam transport, this crosstalk implies parasitic interferences before the recombination step, so that the ``intrinsic'' coherence between the beams, and consequently the resulting interferometric observables, is modified. \citet{2009ApJ...706.1299M} studied for the first time the impact of parasitic interference on the measurement of the complex visibility in classical stellar interferometry. The degradation of the modulus and the phase of the complex visibility depends on two parameters : the residual piston and the crosstalk level - or contamination rate - between the interferometric beams. This degradation may be significant when considering the detection of close-in extrasolar giant planets with the use of differential phase. Here we extend this study to nulling interferometry with a co-axial recombination mode. We describe how this instrumental effect may affect the transmission map of the interferometer, and impact the geometric and instrumental leakages. \\
This article is organized as follows: in Section~\ref{s:theory}, we develop a simple formalism to describe this problem in the case of a Bracewell interferometer. From that, we derive in Section~\ref{s:response} the modified intensity response of the interferometer as a function of the two relevant parameters for this instrumental effect, namely the crosstalk level and the instrumental phase shift between beams \citep{2009ApJ...706.1299M}. In Section~\ref{s:output}, we derive the modified null output of the interferometer and the corresponding geometric and instrumental leakages. Then, in Section~\ref{s:snr}, we detail the impact of crosstalk on the performance of a nulling interferometer, and we describe the results obtained with end-to-end numerical simulations of a classical Bracewell scheme. Finally, in Section~\ref{s:conclusion}, we summarize our work and develop some perspectives on the implementation of this issue in future instrumental studies. 
\section{Theoretical elements}\label{s:theory}
\subsection{Intensity response of a nulling interferometer}
\begin{figure*}
\centering
		\includegraphics[width=100mm,height=55mm]{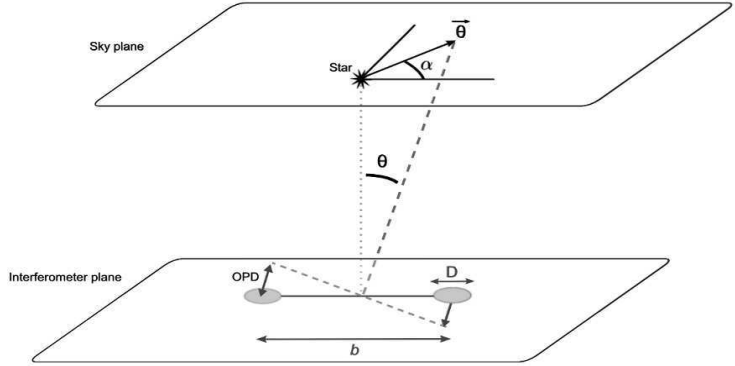}
		\caption{Geometric configuration of the interferometer and the astrophysical source. The two angular coordinates ($\theta$, $\alpha$) give the position in the sky plane. Here, the line-of-sight is assumed to be perpendicular to the interferometer plane.}
		\label{fig:skyscheme}		
\end{figure*}

We summarize the basic equations of nulling interferometry in the most simple case of a two-telescope pupil plane Bracewell interferometer. We keep the same formalism as the one used in \citet{Absil:thesis}. This nulling interferometer is characterized by its intensity response $R_{\lambda}(\boldsymbol{\theta})$, where, $\boldsymbol{\theta}=(\theta,\alpha)$ is the vector giving the coordinates in the sky plane with respect to the optical axis (see Fig.~\ref{fig:skyscheme}). This response or `transmission map' is derived from the addition of the complex amplitudes of the electric field, coming from a point-like source located at the $\boldsymbol{\theta}$ direction in the sky, collected by each telescope and splitted to go either into the destructive or into the constructive output. 
At any point $\boldsymbol{r}$ of the overlapping pupil plane, in the destructive or constructive output, the total electric field is thus given by: 
{\small \begin{equation}
E(\boldsymbol{\theta},\boldsymbol{r})=\Pi(\frac{r}{R})\left[E_1(\boldsymbol{\theta})e^{i\phi_1}+E_2(\boldsymbol{\theta})e^{i\phi_2} \right],
\label{eq:electricfield}
\end{equation}}where $E_1(\boldsymbol{\theta})$ and $E_2(\boldsymbol{\theta})$ are the electric fields corresponding to the telescopes 1 and 2, respectively. $\Pi({\frac{r}{R}})$ is the transmission function of the pupil, $r=\sqrt{x^2+y^2}$ is the distance in the pupil plane, $R$ is the radius of the pupil, and $\phi_{k}$ is the corresponding phase term of the beam $k$. The expression of the complex electric field $E_k(\boldsymbol{\theta})$ is $E_k(\boldsymbol{\theta})=E_ke^{i\frac{2\pi}{\lambda}\boldsymbol{x}_k\boldsymbol{\cdot}\boldsymbol{\theta}}$, where $\boldsymbol{x}_k$ is the position of the telescope $k$. We thus have:
{\small \begin{align}
\nonumber
E(\boldsymbol{\theta},\boldsymbol{r})&=\Pi(\frac{r}{R})\left[ E_1e^{i\left(\frac{2\pi}{\lambda}\boldsymbol{x}_1\boldsymbol{\cdot}\boldsymbol{\theta}+\phi_1\right)}+E_2e^{i\left(\frac{2\pi}{\lambda}\boldsymbol{x}_2\boldsymbol{\cdot}\boldsymbol{\theta}+\phi_2\right)}\right], \\
&=\Pi(\frac{r}{R})r_{\rm \lambda}(\boldsymbol{\theta})\, ,
\label{eq:amplituderesponse}
\end{align}}where $r_{\rm \lambda}$ is defined as the amplitude response of the interferometer. From Eq.~\ref{eq:amplituderesponse}, we can derive the intensity response of the nulling interferometer, noted $R_{\rm \lambda}(\boldsymbol{\theta})$:
{\small \begin{align}
\nonumber
R_{\rm \lambda}(\boldsymbol{\theta})&=r_{\rm \lambda}(\boldsymbol{\theta})r^*_{\rm \lambda}(\boldsymbol{\theta}),\\
\nonumber
&= |E_1|^2+|E_2|^2 \\
&+2E_1E_2\cos\left((\phi_1-\phi_2)+\frac{2\pi}{\lambda}(\boldsymbol{x_1}-\boldsymbol{x_2})\boldsymbol{\cdot}\boldsymbol{\theta}\right) 
\label{eq:intensityresponse}
\end{align}}Assuming that each beam is divided into two equal parts by the beam splitter, and that they were collected by two telescopes of unitary size, we have $|E_1|^2=|E_2|^2=1/2$. Following the notations of Fig.~\ref{fig:skyscheme}, we have $(\boldsymbol{x}_1-\boldsymbol{x}_2)\boldsymbol{\cdot}\boldsymbol{\theta}=b\theta\cos(\alpha)$. Moreover, in the destructive output of the beam splitter, the beams will be $\pi$-shifted, namely $\phi_1-\phi_2=\pi$ radians. The final expression of the intensity response is given by:
{\small \begin{align}
R_{\rm \lambda}(\theta,\alpha) &= 2\sin^2\left(\pi\frac{b\theta}{\lambda}\cos(\alpha)\right)\, ,
\label{eq:response}
\end{align}}where no phase perturbation is taken into account. In the presence of phase perturbation, occuring for instance because of vibrations inside the interferometer arms or imperfect piston correction by a fringe tracker, the intensity response becomes: 
{\small \begin{align}
R_{\rm \lambda}(\theta,\alpha) &= 2\sin^2\left(\pi\frac{b\theta}{\lambda}\cos(\alpha)+\frac{\delta\phi(t)}{2}\right)\, ,
\label{eq:responseinstrumental}
\end{align}}
where $\delta \phi(t)$ is the variable optical path delay between the beams of the interferometer.

\subsection{Parasitic Interference Model}
To create a model of parasitic interference, we consider the same formalism as the one described above, in the case of a two-telescope pupil plane Bracewell interferometer. 
However, because of multiple parasitic reflections inside transmitting optics and/or diffraction effects associated with beam propagation along finite size optics, a coherent crosstalk may occur between a beam and itself, and/or between each of the individual beams (mutual beam contamination). This crosstalk, occuring during the beam transport, implies parasitic interferences. They will modify the `intrinsic' coherence between beams and then the intensity response of the nulling interferometer, in the destructive and constructive outputs. To model it, we define $\epsilon_1$ (resp. $\epsilon_2$) as the main fraction of $E_1(\boldsymbol{\theta})$ (resp.~$E_2(\boldsymbol{\theta})$) propagating along the path 1 (resp. 2), and hereafter considered as `the primary beam'. The two causes of crosstalk are expected to produce the same effect, namely splitting and phase-shifting some light from the main beam. Therefore, we define $\epsilon'_1$ as the small fraction of $E_1(\boldsymbol{\theta})$, either reflected twice inside the transmitting optics and still following the path 1, or having contaminated $E_2(\boldsymbol{\theta})$ and following the path 2 (see Fig.~\ref{fig:crosstalk}). It is hereafter considered as the `secondary beam'. $\epsilon'_2$ corresponds to the small parasitic fraction associated to $E_2(\boldsymbol{\theta})$. All the parasitic reflections and/or beam contaminations occuring inside the instrument produce a resulting modified pattern in the overlapping pupil plane of each interferometric output. 
At any point $\boldsymbol{r}$ of this plane, the total electric field is thus given by : 
\begin{align}
\nonumber
E(\boldsymbol{\theta},\boldsymbol{r})&=\Pi(r/R)\left[\epsilon_1 E_1(\boldsymbol{\theta})e^{i\phi_1} +\epsilon_1' E_1(\boldsymbol{\theta})e^{i\phi_1'}\right.\\ 
&\left.+\epsilon_2 E_2(\boldsymbol{\theta})e^{i\phi_2}+\epsilon_2' E_2(\boldsymbol{\theta})e^{i\phi_2'} \right],
\label{eq:electricfieldpar}
\end{align}where $\phi_{k}$ and $\phi_{k}'$ are the corresponding phase terms of the main part and the secondary - or parasitic - part of the beam $k$, respectively. We implicitely assume here that the crosstalk is fully coherent.  

\section{Modified intensity response}\label{s:response}
\subsection{General expression}
From Eq.~\ref{eq:electricfieldpar}, we derive the modified intensity response of the nulling interferometer, noted $\tilde{R}_{\rm \lambda}(\boldsymbol{\theta})$, in the presence of crosstalk. Using the same notations as in Section~2.1, we thus have :
\begin{figure*}
\centering
		\includegraphics[width=100mm,height=40mm]{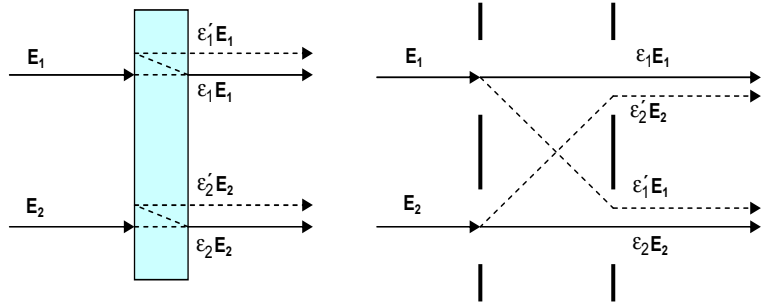}
		\caption{Simple scheme describing the different possibilities of beam contamination (or crosstalk). This beam contamination can be produced by a parasitic reflection inside transmitting optics (left) or by a beam mixing (right). $\epsilon_1 E_1$ and $\epsilon_2 E_2$ represent the main beams (in electrical field), while $\epsilon'_1 E_1$ ans $\epsilon'_2 E_2$ corresponds to the secondary beams reflected inside the optics and/or going though the wrong paths (for example due to diffraction or beam mixing).}
		\label{fig:crosstalk}		
\end{figure*}{\small \begin{align}
\nonumber
E(\boldsymbol{\theta},\boldsymbol{r})&=\Pi(r/R)\left[\epsilon_1 E_1e^{i\left(\frac{2\pi}{\lambda}\boldsymbol{x}_1\boldsymbol{\cdot}\boldsymbol{\theta}+\phi_1\right)}+\epsilon_2 E_2e^{i\left(\frac{2\pi}{\lambda}\boldsymbol{x}_2\boldsymbol{\cdot}\boldsymbol{\theta}+\phi_2\right)} \right. \\
\nonumber
&\left.+\epsilon_1'E_1e^{i\left(\frac{2\pi}{\lambda}\boldsymbol{x}_1\boldsymbol{\cdot}\boldsymbol{\theta}+\phi_1'\right)}+\epsilon_2'E_2e^{i\left(\frac{2\pi}{\lambda}\boldsymbol{x}_2\boldsymbol{\cdot}\boldsymbol{\theta}+\phi_2'\right)}\right], \\
&=\Pi(r/R)\tilde{r}_{\rm \lambda}(\boldsymbol{\theta}).
\label{eq:amplituderesponseperturbed}
\end{align}}Then we calculate the modified intensity response of the interferometer :
{\small \begin{align}
\nonumber
\tilde{R}_{\rm \lambda}(\boldsymbol{\theta})&=\tilde{r}_{\lambda}(\boldsymbol{\theta})\tilde{r}_{\lambda}^*(\boldsymbol{\theta}),\\
\nonumber
&= \epsilon_1^2|E_1|^2+\epsilon_2^2|E_2|^2 \\
\nonumber
&+2\epsilon_1\epsilon_2E_1E_2\cos\left(\frac{2\pi}{\lambda}(\boldsymbol{x_1}-\boldsymbol{x_2})\boldsymbol{\cdot}\boldsymbol{\theta}+(\phi_1-\phi_2)\right) \\
\nonumber
&+2\epsilon_1\epsilon'_1|E_1|^2\cos(\phi_1-\phi_1')+2\epsilon_2\epsilon'_2|E_2|^2\cos(\phi_2-\phi_2') \\
\nonumber
&+2\epsilon_1\epsilon'_2E_1E_2\cos\left(\frac{2\pi}{\lambda}(\boldsymbol{x_1}-\boldsymbol{x_2})\boldsymbol{\cdot}\boldsymbol{\theta}+(\phi_1-\phi_2')\right) \\
\nonumber
&+2\epsilon_2\epsilon'_1E_1E_2\cos\left(\frac{2\pi}{\lambda}(\boldsymbol{x_1}-\boldsymbol{x_2})\boldsymbol{\cdot}\boldsymbol{\theta}+(\phi_1'-\phi_2)\right)\\
\nonumber
&+\epsilon_1'^2|E_1|^2+\epsilon_2'^2|E_2|^2 \\
&+2\epsilon_1'\epsilon_2'E_1E_2\cos\left(\frac{2\pi}{\lambda}(\boldsymbol{x_1}-\boldsymbol{x_2})\boldsymbol{\cdot}\boldsymbol{\theta}+(\phi_1'-\phi_2')\right)
\label{eq:intensityresponseperturbed}
\end{align}}As in Section 2.1, we assume that the corresponding intensity collected by each of the two telescopes is normalised, and that each corresponding beam is divided into two equal parts by the balanced beam splitter, which implies {\small $|E_1|^2=|E_2|^2=1/2$}. We also consider a differential effect between the beams in the contamination process, namely: $\epsilon_2=\epsilon_1+\Delta\epsilon$ and $\epsilon_2'=\epsilon_1'+\Delta\epsilon'$. The general expression of the modified intensity response then becomes :
{\small \begin{align}
\nonumber
\tilde{R}_{\rm \lambda}(\theta,\alpha) &= \epsilon_1^2\left[1+\cos\left(2\pi\frac{b\theta}{\lambda}\cos(\alpha)+(\phi_1-\phi_2)\right)\right] \\
\nonumber
&+\epsilon_1'^2\left[1+\cos\left(2\pi\frac{b\theta}{\lambda}\cos(\alpha)+(\phi_1'-\phi_2')\right)\right] \\
\nonumber
&+\epsilon_1\epsilon_1'\left[\vphantom{\cos(2\pi\frac{b\theta}{\lambda}\cos(\alpha)+(\phi_1-\phi_2+\Delta\phi_2))}\cos(\phi_1-\phi_1')+\cos(\phi_2-\phi_2')\right. \\
\nonumber
&+\cos\left(2\pi\frac{b\theta}{\lambda}\cos(\alpha)+(\phi_1-\phi_2')\right) \\
\nonumber
&\left.+\cos\left(2\pi\frac{b\theta}{\lambda}\cos(\alpha)+(\phi_1'-\phi_2)\right)\right] \\
\nonumber
&+\epsilon_1\Delta\epsilon+\frac{\Delta\epsilon^2}{2}+\epsilon'_1\Delta\epsilon'+\frac{\Delta\epsilon'^2}{2} \\
\nonumber
&+(\epsilon_1\Delta\epsilon'+\Delta\epsilon_1'+\Delta\epsilon\Delta\epsilon')\cos(\phi_2-\phi_2') \\
\nonumber
&+\epsilon_1\Delta\epsilon\cos\left(2\pi\frac{b\theta}{\lambda}\cos(\alpha)+(\phi_1-\phi_2)\right) \\
\nonumber
&+\epsilon_1\Delta\epsilon'\cos\left(2\pi\frac{b\theta}{\lambda}\cos(\alpha)+(\phi_1-\phi_2')\right) \\
&+\epsilon_1'\Delta\epsilon\cos\left(2\pi\frac{b\theta}{\lambda}\cos(\alpha)+(\phi_1'-\phi_2)\right). 
\label{eq:responseperturbed}
\end{align}}Assuming a coherent normalization between the primary and secondary (or parasitic) beams, namely $\epsilon_1+\epsilon_1'=1$ and $\epsilon_2+\epsilon_2'=1$, we have $\Delta\epsilon=-\Delta\epsilon'$. Then, assuming that each secondary beam is phase-shifted with respect to its primary beam, namely $\phi_k-\phi_k'=\Delta \phi_k$, we can thus rewrite Eq.~\ref{eq:responseperturbed} as:
{\small \begin{equation}
\tilde{R}_{\rm \lambda}(\theta,\alpha)=\tilde{R}_{\rm \lambda,bal}(\theta,\alpha)+\tilde{R}_{\rm \lambda,unbal}(\theta,\alpha),
\end{equation}
}
with:
{\small \begin{align}
\nonumber
\tilde{R}_{\rm \lambda,bal}(\theta,\alpha)&= \epsilon_1^2\left[1+\cos\left(2\pi\frac{b\theta}{\lambda}\cos(\alpha)+(\phi_1-\phi_2)\right)\right] \\
\nonumber
&+\epsilon_1'^2\left[1+\cos\left(2\pi\frac{b\theta}{\lambda}\cos(\alpha) \right. \right.\\
\nonumber
&\left.\left.+(\phi_1-\phi_2+\Delta\phi_2-\Delta\phi_1)\vphantom{\cos(2\pi\frac{b\theta}{\lambda}\cos(\alpha)+(\phi_1-\phi_2+\Delta\phi_2))}\right)\vphantom{\cos(2\pi\frac{b\theta}{\lambda}\cos(\alpha)+(\phi_1-\phi_2+\Delta\phi_2))}\right] \\
\nonumber
&+\epsilon_1\epsilon_1'\left[\vphantom{\cos(2\pi\frac{b\theta}{\lambda}\cos(\alpha)+(\phi_1-\phi_2+\Delta\phi_2))}\cos(\Delta\phi_1)+\cos(\Delta\phi_2)\right. \\
\nonumber
&+\cos\left(2\pi\frac{b\theta}{\lambda}\cos(\alpha)+(\phi_1-\phi_2-\Delta\phi_1)\right) \\
&\left.+\cos\left(2\pi\frac{b\theta}{\lambda}\cos(\alpha)+(\phi_1-\phi_2+\Delta\phi_2)\right)\right],
\label{eq:balancedresponse}
\end{align}}
and
{\small \begin{align}
\nonumber
\tilde{R}_{\rm \lambda,unbal}(\theta,\alpha) &= \Delta\epsilon(\epsilon_1-\epsilon_1')+\Delta\epsilon^2-\Delta\epsilon^2\cos(\Delta\phi_2) \\
\nonumber
&+\epsilon_1\Delta\epsilon\left[\cos\left(2\pi\frac{b\theta}{\lambda}\cos(\alpha)+(\phi_1-\phi_2)\right) \right. \\
\nonumber
&-\cos(\Delta\phi_2) \\
\nonumber
&\left.-\cos\left(2\pi\frac{b\theta}{\lambda}\cos(\alpha)+(\phi_1-\phi_2+\Delta\phi_2)\right)\right] \\
\nonumber
&+\epsilon_1'\Delta\epsilon\left[-\cos\left(2\pi\frac{b\theta}{\lambda}\cos(\alpha)\right. \right. \\
\nonumber
&\left.+(\phi_1-\phi_2+\Delta\phi_2-\Delta\phi_1)\vphantom{\cos(2\pi\frac{b\theta}{\lambda}\cos(\alpha)+(\phi_1-\phi_2+\Delta\phi_2))}\right) +\cos(\Delta\phi_2) \\
&\left.+\cos\left(2\pi\frac{b\theta}{\lambda}\cos(\alpha)+(\phi_1-\phi_2-\Delta\phi_1)\right)\right].
\label{eq:unbalancedresponse}
\end{align}}
It thus appears that the expression of the modified intensity response can be separated in two terms, indicated by the subscripts 'bal' and 'unbal' that stand for balanced and unbalanced, respectively. Indeed, $\tilde{R}_{\rm \lambda,bal}(\theta,\alpha)$ does not depend on $\Delta\epsilon$, the crosstalk imbalance, while $\tilde{R}_{\rm \lambda,unbal}(\theta,\alpha)$ does.\\ 
Following the basic principle of nulling interferometry, a relative $\pi$-phase shift is applied between the two beams before recombination, namely $\phi_1-\phi_2=\pi$ radians. In addition, an instrumental phase perturbation may occur for instance because of vibrations inside the interferometer arms or imperfect piston correction by a fringe tracker. Therefore, a variable instrumental phase shift $\delta \phi(t)$ is also affecting the beams in the interferometric arms. In the destructive output of the beam splitter, we thus have : $\phi_1-\phi_2=\pi+\delta\phi(t)$. 
The two terms of the modified intensity response can thus be written as :
{\small \begin{align}
\nonumber
\tilde{R}_{\rm \lambda,bal}(\theta,\alpha)&= \epsilon_1^2\left[1-\cos\left(2\pi\frac{b\theta}{\lambda}\cos(\alpha)+\delta \phi(t)\right)\right] \\
\nonumber
&+\epsilon_1'^2\left[1-\cos\left(2\pi\frac{b\theta}{\lambda}\cos(\alpha)+\delta \phi(t) \right. \right.\\
\nonumber
&\left.\left.+(\Delta\phi_2-\Delta\phi_1)\vphantom{\cos\left(2\pi\frac{b\theta}{\lambda}\cos(\alpha)+\delta \phi(t)\right)}\right)\vphantom{\cos\left(2\pi\frac{b\theta}{\lambda}\cos(\alpha)+\delta \phi(t)\right)}\right] \\
\nonumber
&+\epsilon_1\epsilon_1'\left[\vphantom{\cos(2\pi\frac{b\theta}{\lambda}\cos(\alpha)+(\phi_1-\phi_2+\Delta\phi_2))}\cos(\Delta\phi_1)+\cos(\Delta\phi_2)\right. \\
\nonumber
&-\cos\left(2\pi\frac{b\theta}{\lambda}\cos(\alpha)+\delta \phi(t)-\Delta\phi_1)\right) \\
&\left.-\cos\left(2\pi\frac{b\theta}{\lambda}\cos(\alpha)+\delta \phi(t)+\Delta\phi_2)\right)\right],
\label{eq:balancedresponsedest}
\end{align}}
and
{\small \begin{align}
\nonumber
\tilde{R}_{\rm \lambda,unbal}(\theta,\alpha) &= \Delta\epsilon(\epsilon_1-\epsilon_1')+\Delta\epsilon^2-\Delta\epsilon^2\cos(\Delta\phi_2) \\
\nonumber
&+\epsilon_1\Delta\epsilon\left[-\cos\left(2\pi\frac{b\theta}{\lambda}\cos(\alpha)+\delta \phi(t)\right) \right. \\
\nonumber
&\left.-\cos(\Delta\phi_2)+\cos\left(2\pi\frac{b\theta}{\lambda}\cos(\alpha)+\delta \phi(t)+\Delta\phi_2\right)\right] \\
\nonumber
&+\epsilon_1'\Delta\epsilon\left[\cos\left(2\pi\frac{b\theta}{\lambda}\cos(\alpha)+\delta \phi(t)+\Delta\phi_2-\Delta\phi_1\right) \right. \\
&\left.+\cos(\Delta\phi2)-\cos\left(2\pi\frac{b\theta}{\lambda}\cos(\alpha)+\delta \phi(t)-\Delta\phi_1\right)\right].
\label{eq:unbalancedresponsedest}
\end{align}}
In the following, we describe the dependency of the modified intensity response on the crosstalk levels or contamination rates ($\epsilon_1'^2$ and $\epsilon_2'^2$) and the phase shift between the primary and secondary (or parasitic) beams ($\Delta\phi_1$ and $\Delta\phi_2$). We first consider the simple case of an equal crosstalk level between both beams, namely $\epsilon_1'=\epsilon_2'=\epsilon'$). Then, we consider in addition the impact of a differential crosstalk effect ($\Delta\epsilon\neq0$) on the intensity response, through the term $\tilde{R}_{\rm \lambda,unbal}(\theta,\alpha)$. Finally, we briefly address the impact of a finite bandwidth on the coherent crosstalk perturbation.

\subsection{equal crosstalk}
\begin{figure}
\centering
		\includegraphics[width=80mm,height=55mm]{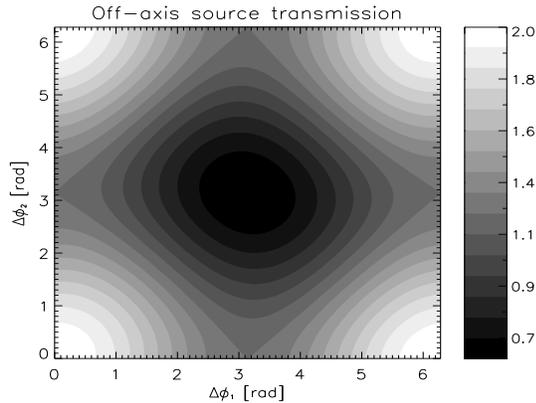}
		\caption{
		Evolution of the value of $\tilde{R}_{\lambda}(\theta_{\rm pl},\alpha_{\rm pl})$, at 10~$\mu$m, as a function of the phase shifts $\Delta \phi_1$ and $\Delta \phi_2$, between the primary and secondary beams. The angular position of the off-axis source is assumed to be ($\theta_{\rm pl},\alpha_{\rm pl}$), with $\sin^2\left(\pi\frac{b\theta_{\rm pl}}{\lambda}\cos(\alpha_{\rm pl})\right)=1$. We here consider $\epsilon=0.8$ and $\epsilon'=0.2$, and no instrumental phase shift.}
		\label{fig:evoltrans}		
\end{figure}

Assuming $\epsilon_1'=\epsilon_2'=\epsilon'$, the modified intensity response can be simplified as:
{\small \begin{align}
\nonumber
\tilde{R}_{\rm \lambda}(\theta,\alpha) &= \epsilon^2\left[1-\cos\left(2\pi\frac{b\theta}{\lambda}\cos(\alpha)+\delta \phi(t)\right)\right] \\
\nonumber
&+\epsilon\epsilon'\left[\vphantom{\frac{2\pi}{\lambda}}\cos(\Delta \phi_1)+\cos(\Delta \phi_2)\right. \\
\nonumber
&-\cos\left(2\pi\frac{b\theta}{\lambda}\cos(\alpha)+\delta \phi(t)-\Delta \phi_1\right)\\
\nonumber
&\left.-\cos\left(2\pi\frac{b\theta}{\lambda}\cos(\alpha)+\delta \phi(t)+\Delta \phi_2\right)\right] \\
&+\epsilon'^2\left[1-\cos\right(2\pi\frac{b\theta}{\lambda}\cos(\alpha)+\delta \phi(t)+(\Delta \phi_2-\Delta \phi_1)\left)\right].
\label{eq:generalresponsedest-equal}
\end{align}}
From this expression, we clearly see that the effect of the beam contamination on the transmission map depends on the value of $\Delta \phi_k$. According to Eq.~\ref{eq:generalresponsedest-equal}, we show in Fig.\ref{fig:evoltrans} the evolution of the transmission of an off-axis source as a function of $\Delta \phi_1$ and $\Delta \phi_2$ for a given amount of crosstalk (i.e., $\epsilon=0.8$ and $\epsilon'=0.2$). The angular position of the off-axis source is assumed to be ($\theta_{\rm pl},\alpha_{\rm pl}$), with $\sin^2\left(\pi\frac{b\theta_{\rm pl}}{\lambda}\cos(\alpha_{\rm pl})\right)=1$. It clearly appears that the transmission is maximum when no phase shift is produced during the contamination process, independently of the crosstalk level. In this case, namely $\Delta \phi_1=\Delta \phi_2=0$, the transmission map indeed becomes : $R(\lambda,\boldsymbol{\theta})=2(\epsilon+\epsilon')^2\sin^2\left(\pi\frac{b\theta}{\lambda}\cos(\alpha)+\frac{\delta\phi(t)}{2}\right)$. Here, the primary and secondary parts of each beam remain co-phased in the overlapping pupil plane and do not imply a modification of the intensity response. This is represented by the factor $(\epsilon+\epsilon')^2$, which is equal to 1.       
In contrast, as shown in Fig.~\ref{fig:evoltrans}, the transmission of the off-axis source is minimized when the parasitic beams are both $\pi$ phase-shifted. In this case, the modified intensity response becomes :
{\small \begin{equation}
\tilde{R}_{\lambda}(\theta,\alpha)=2(\epsilon-\epsilon')^2\sin^2\left(\pi\frac{b\theta}{\lambda}\cos(\alpha)+\frac{\delta\phi(t)}{2}\right).
\label{eq:nophaseresponsepi}
\end{equation}}In this case, we can clearly notice a decrease of the level of the transmission map in the destructive output of the interferometer, for every angular position, by a factor $(\epsilon-\epsilon')^2$.
\begin{figure}
\centering
		\includegraphics[width=75mm,height=55mm]{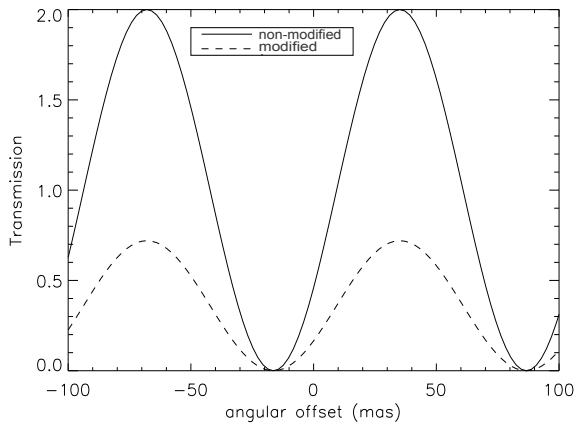}
		\caption{Plot of the 1D transmission map (or intensity response) of a Bracewell nulling interferometer having a baseline length of 20~m and observing at 10~$\mu$m. We consider the non-modified case (without crosstalk), i.e. $R_{\lambda}(\theta,\alpha)$ (solid line), and the modified case (with crosstalk), i.e. $\tilde{R}_{\lambda}(\theta,\alpha)$  (dashed line). In addition, we assume $\epsilon=0.8$ and $\epsilon'=0.2$, $\Delta\phi_1=\Delta\phi_2=\pi$, and an instrumental phase shift of 1 radian.}
		\label{fig:evoltrans2}		
\end{figure}
This is illustrated in Fig.~\ref{fig:evoltrans2}, where we show the transmission map of a Bracewell interferometer with and without crosstalk (see dashed and solid curves respectively). As a consequence of this overall decrease, the transmission close to the centre of the field-of-view is lower than in the nominal case without coherent crosstalk. The flux coming from the on-axis stellar source is thus better rejected, which implies less stellar leakage, while the flux of the off-axis source is less transmitted. This result is not surprising if we refer to the description of \citet{2009ApJ...706.1299M}. In the case of crosstalk occuring in a classical interferometer, parasitic interference creates two other fringe patterns in addition to the intrinsic fringe pattern due to the astrophysical source. One of them can be assimilated to a Young-like fringe pattern, which is created by two sources, possibly coherent, represented by the main part of the beam (noted $\epsilon$) and the small contribution having contaminated the other path (noted $\epsilon'$). This pattern artificially increases the proportion of coherent flux in the resulting interference pattern, since a part of the incoherent flux of the source was actually used to form these Young-like fringes. In our case, the apparent better rejection of the on-axis source flux is thus accompanied by a lower transmission of the flux of the off-axis source (see Fig.\ref{fig:evoltrans2}). The signal coded in the modified interferogram and forming the transmission map, does not only refer to the frequency information of the astrophysical source at $b/\lambda$, but also to a contribution of the lower spatial frequencies originating from the transfer function of the individual apertures. 

\subsection{Differential crosstalk}
\begin{figure*}
\centering
		\includegraphics[width=80mm,height=65mm]{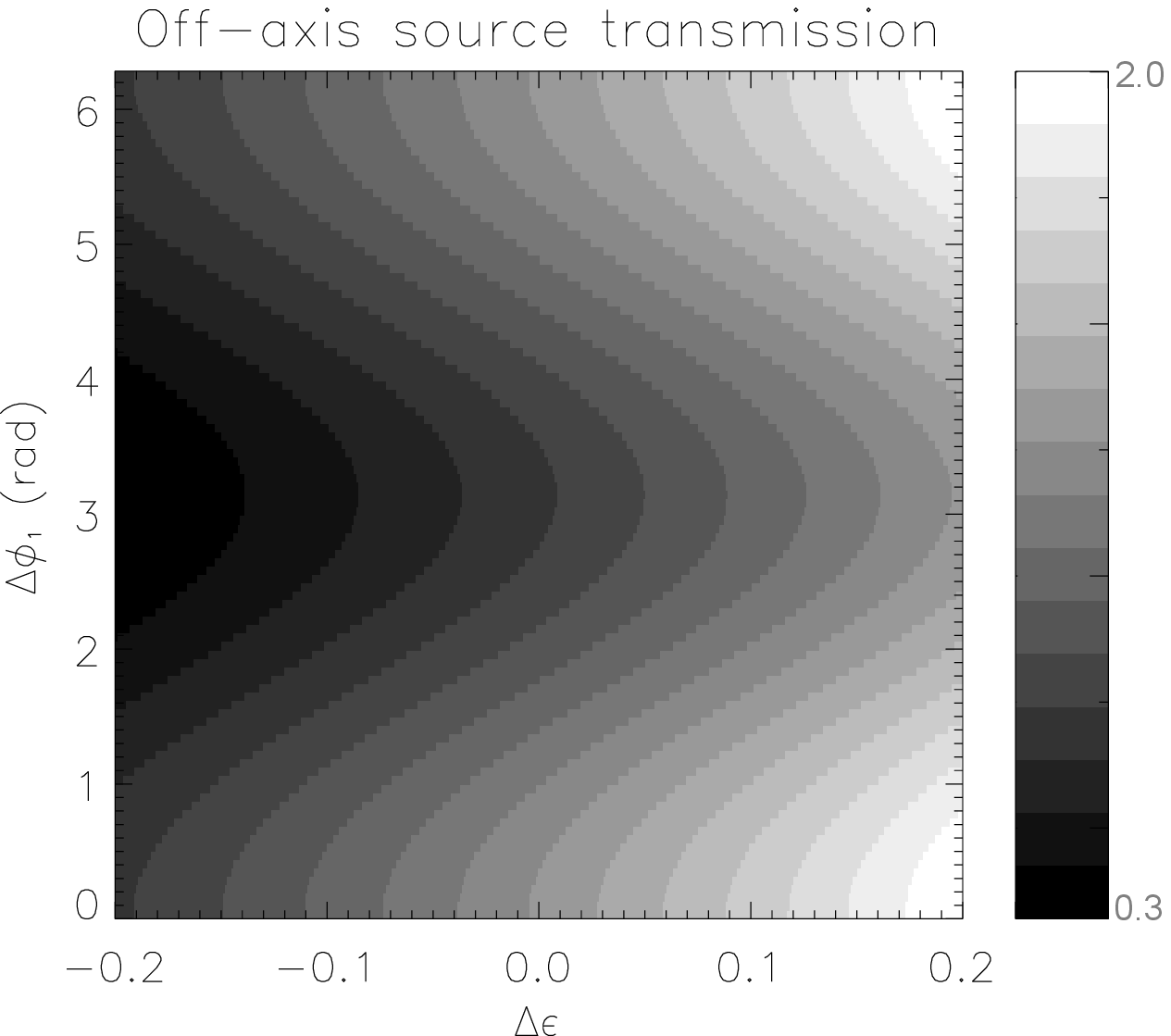} \hfill
		\includegraphics[width=80mm,height=65mm]{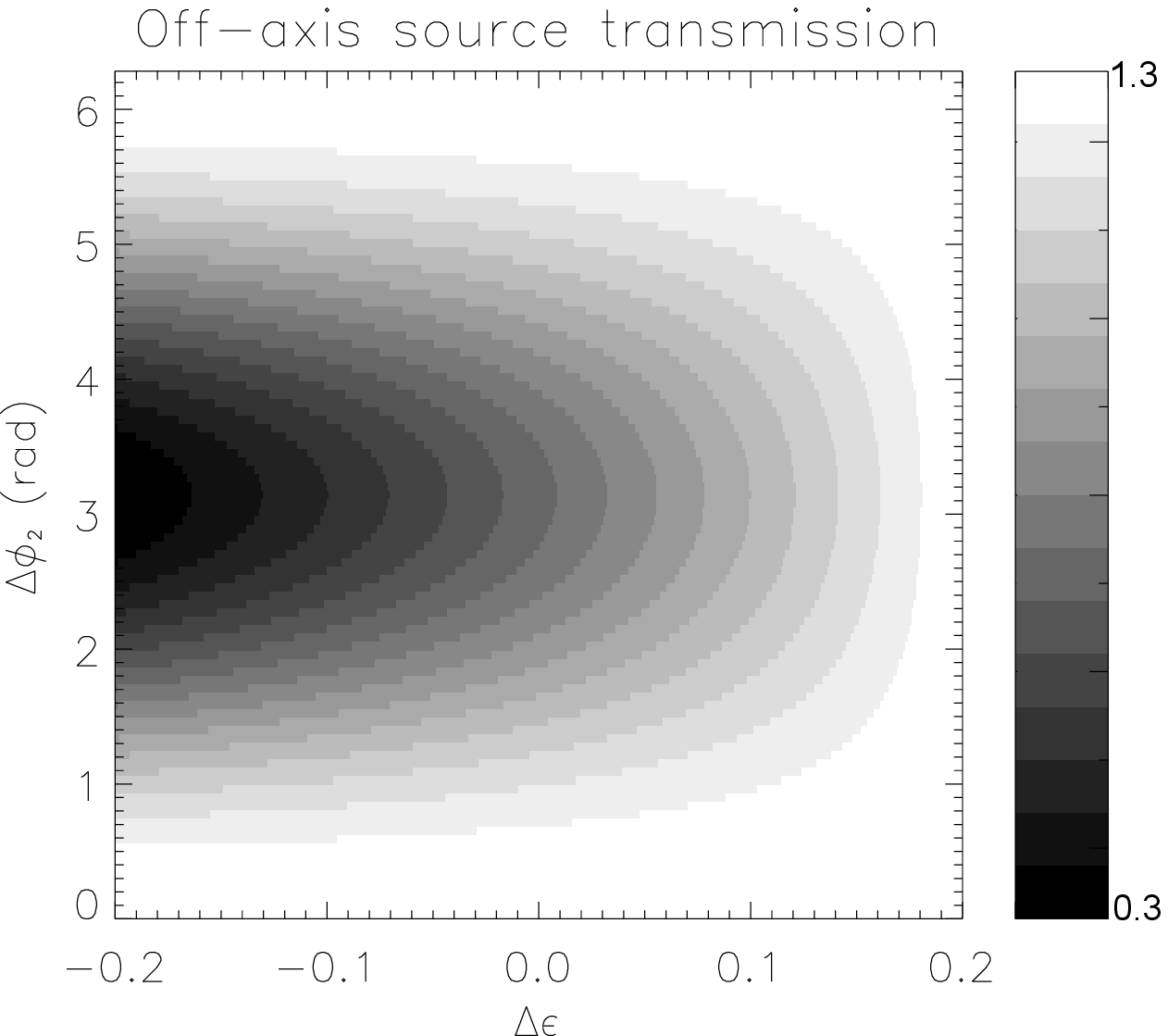}
		\caption{
		{\itshape Left}: Modified transmission of an off-axis source, $\tilde{R}_{\lambda}(\theta_{\rm off},\alpha_{\rm off})$,  at 10~$\mu$m, as a function of the differential crosstalk $\Delta\epsilon$ and the phase shift $\Delta \phi_1$; assuming $\Delta \phi_2=\pi$. This off-axis source is assumed to be on a maximum of the ideal transmission map, namely $\sin^2\left(\pi\frac{b\theta_{\rm pl}}{\lambda}\cos(\alpha_{\rm pl})\right)=1$. In addition, we consider $\epsilon=0.8$, $\epsilon'=0.2$, $\Delta\epsilon=0.1$, and no instrumental phase shift. {\itshape Right}: Same as {\itshape Left} but plotted as a function of $\Delta\epsilon$ and $\Delta\phi_2$, and assuming $\Delta \phi_1=\pi$.}
		\label{fig:evoltrans3}		
\end{figure*}

Assuming differential effects for the beam contamination process, namely $\Delta\epsilon\neq0$, we consider here the modified intensity response composed of the two terms $\tilde{R}_{\rm \lambda,bal}$ and $\tilde{R}_{\rm \lambda,unbal}$ (see Eqs.~\ref{eq:balancedresponsedest} and \ref{eq:unbalancedresponsedest}). The effect of differential crosstalk is represented in Fig.~\ref{fig:evoltrans3}, showing the off-axis source transmission with respect to $\Delta\epsilon$ and $\Delta\phi_1$, assuming $\Delta\phi_2=\pi$ (left), and with respect to  $\Delta\epsilon$ and $\Delta\phi_2$, assuming $\Delta\phi_1=\pi$ (right). In the framework of our parasitic interference model, a positive differential crosstalk will improve the off-axis transmission compared to the `equal crosstalk' case ($\Delta\epsilon=0$). By comparing the left and right panels of Fig.~\ref{fig:evoltrans3}, it also appears that a differential crosstalk causes an asymmetry in the effect of $\Delta\phi_1$ and $\Delta\phi_2$ on the off-axis transmission. Choosing $\Delta\phi_1=\pi$ rather than $\Delta\phi_2=\pi$ implies a lower maximum for the off-axis transmission value. However, the minimum value of the off-axis transmission remains the same in both cases and is still associated to $\Delta\phi_1=\Delta\phi_2=\pi$. 
In this case, the modified intensity response simplifies as:
{\small \begin{equation}
\tilde{R}_{\lambda}(\theta,\alpha)=2(\epsilon_1-\epsilon_1'+\Delta\epsilon)^2\sin^2\left(\pi\frac{b\theta}{\lambda}\cos(\alpha)+\frac{\delta\phi(t)}{2}\right)
\label{eq:nophaseresponsepi-diff}
\end{equation}}
On the contrary, if no phase shift occurs during the contamination process, equal and differential crosstalk effects cancel out and we retrieve the intrinsic expression of the intensity response, only affected by the instrumental phase shift $\delta\phi(t)$: $\tilde{R}_{\lambda}(\theta,\alpha)=(\epsilon_1+\epsilon_1')^2\left[ 1-\cos\left(2\pi\frac{b\theta}{\lambda}\cos(\alpha)+\delta\phi(t)\right)\right]=2\sin^2\left(\pi\frac{b\theta}{\lambda}\cos(\alpha)+\frac{\delta\phi(t)}{2}\right)$.\\
Therefore we clearly see that the differential crosstalk will impact the level of the transmission map, depending on the value and the sign of $\Delta\epsilon$.\\
In the next Section, we derive the crosstalk-affected null output and assess the impact on the associated signal-to-noise ratio. In order to highlight the general effect of crosstalk on those quantities, we will assume an equal crosstalk between the beams ($\Delta\epsilon=0$).

\subsection{Finite bandwidth}\label{s:bandwidth}
The previous discussion assumes that light is purely monochromatic so that the coherence time is infinitely long. In practice, interferometric observations are generally obtained with a finite bandwidth so that the coherence is only ensured within a given coherence length, called $L$ in the following. When the optical path differences associated with $\Delta\phi_k$ are longer than $L$, the parasitic beams will not interfere anymore with the main beams. In addition, if the optical path difference between the parasitic beams is also longer than the coherence length, the crosstalk is then equivalent to a classical intensity mismatch effect described in detail elsewhere \citep[e.g.,][]{2004ApOpt..43.6100L}. However, interference might still occur between the parasitic beams if they remain co-phased as it is likely to be the case for an interferometer designed in a symmetric way. This effect is in second order in $\epsilon'$, and the modification of the intensity response can still be described by Eq.~\ref{eq:balancedresponsedest} or Eq.~\ref{eq:unbalancedresponsedest}.

\section{Modified null output} \label{s:output}
The intensity response projected on the plane of the sky forms the transmission map of the interferometer. We here assume that the final detection is performed in an image plane. Therefore, a single-dish telescope image is formed, except that the contribution of each source is affected by the intensity response of the interferometer, depending on their location in the image plane. No fringe is formed nor recorded, and the final output generally consists of the total intensity in the diffraction limited field of view, namely the size of the Airy pattern. Following the mathematical description of \citet{Absil:thesis} , this final output or ``null'' then writes :
{\small \begin{align}
\nonumber
N(\lambda)&=T(\lambda)\int\int{\left[B_{\rm sky}(\lambda,\theta,\alpha)\tilde{R}_{\rm \lambda}(\theta,\alpha)\right.} \\
&\left.+B_{\rm bckg}(\lambda,\theta,\alpha,t)\vphantom{\int{B_{\rm sky}(\lambda,\theta,\alpha)\tilde{R}_{\rm \lambda}(\theta,\alpha)}}\right]P(\theta,\alpha)\theta d\theta d\alpha,
\label{eq:flux}
\end{align}}with $T(\lambda)$ the wavelength-dependent total transmission of the interferometer, $B_{\rm sky}(\lambda,\theta,\alpha)$ the brightness distribution of the source in the diffraction-limited field of view, $B_{\rm bckg}(\lambda,\theta,\alpha,t)$ the time-dependent brightness of the incoherent background emission (sky thermal emission, telescope, optical train), $P(\theta,\alpha)$ the point spread function of a single telescope, and finally $\tilde{R}_{\rm \lambda}(\theta,\alpha)$ the transmission map of the interferometer affected by parasitic interference. Here, we consider a magnification factor of 1, so that we keep the same angular coordinates in the focal plane, namely ($\theta,\alpha$). 

Let us consider the observation of an extended source, composed of a star partially resolved by the interferometer and a secondary component (planet, exozodiacal disc, ...). The corresponding brightness distribution on the sky is :
{\small \begin{equation}
B_{\rm sky}(\lambda,\theta,\alpha)=B_*(\lambda)\Pi(\frac{\theta}{\theta_*})+B_{\rm off}(\lambda)f(\theta,\alpha).
\label{eq:brightdistrib}
\end{equation}}
$\Pi(\frac{\theta}{\theta_*})$ is the top-hat function, which is equal to 1 in the angular domain [0, $\theta_*$], where $\theta_*$ is the angular diameter of the stellar photosphere, and equal to 0 outside. The stellar brightness per square meter per steradian per spectral bandwidth $B_*(\lambda)$ is considered to be constant over the stellar surface (uniform disc approximation). $B_{\rm off}(\lambda)$ is the brightness of the off-axis source, while $f(\theta,\alpha)$ is its distribution on the sky.\\
To illustrate how the null output will be pertubed by the crosstalk effect, we hereafter consider the case of $\Delta\phi_1=\Delta\phi_2=\pi$, so that: {\small $\tilde{R}_{\lambda}(\theta,\alpha)=2(\epsilon-\epsilon')^2\sin^2\left(\pi\frac{b\theta}{\lambda}\cos(\alpha)+\frac{\delta\phi}{2}\right)$}. Then, assuming that the stellar angular diameter is small compared to the fringe spacing ($\theta_* \ll \frac{\lambda}{b}$), and that $\delta\phi(t)\ll1$, we can simplify the expression of the modified transmission map in the angular domain $[0,\theta_*]$, hereafter noted $\tilde{R}_{\theta_*}(\lambda,\theta,\alpha)$ as :
{\small \begin{equation}
\tilde{R}_{\theta_*}(\lambda,\theta,\alpha)\approx 2(\epsilon-\epsilon')^2\left(\pi\frac{b\theta}{\lambda}\cos(\alpha)+\frac{\delta\phi(t)}{2}\right)^2.
\label{eq:approxresponse}
\end{equation}}Provided that the finite stellar photosphere of angular radius $\theta_*$ is almost unresolved by one single telescope (and assuming that is it also the case for the off-axis source), we can assume $P(\theta)=\frac{2J_1(\pi\theta D/\lambda)}{\pi\theta D/\lambda)}\approx1$ for both sources.  The final output can then be written as : 
{\small \begin{align}
\nonumber
N(\lambda) &\approx T(\lambda)B_*(\lambda)\int_0^{2\pi}\int_0^{\theta_*}{\tilde{R}_{\theta_*}(\lambda,\theta,\alpha)\theta d\theta d\alpha} \\
\nonumber
&+T(\lambda)B_{\rm off}(\lambda)\int\int{\tilde{R}_{\lambda}(\theta,\alpha)f(\theta,\alpha) \theta d\theta d\alpha}\\
\nonumber
&+T(\lambda)\int\int{B_{\rm bckg}(\lambda,\theta,\alpha,t) \theta d\theta d\alpha}. \\
\nonumber
N(\lambda) &\approx N_*(\lambda)+B_{\rm off}(\lambda)\int\int{\tilde{R}_{\lambda}(\theta,\alpha)f(\theta,\alpha) \theta d\theta d\alpha}\\
&+\int\int{B_{\rm bckg}(\lambda,\theta,\alpha,t) \theta d\theta d\alpha}.
\label{eq:finalnull}
\end{align}}This modified null output contains, in addition to the off-axis source and background signals, the stellar leakage contribution noted $N_*(\lambda)$. We clearly see that every source contributing to the null output, except the background emission, is affected by the parasitic interference (or crosstalk) effect through the modified intensity response $\tilde{R}(\lambda,\theta,\alpha)$. In this case, the expression of the stellar leakage is: 
{\small \begin{align}
N_*(\lambda) &\approx T(\lambda)B_*(\lambda)(\epsilon-\epsilon')^2\left[\frac{\pi^3 b^2\theta_*^4}{2\lambda^2}+\frac{\delta\phi^2(t)}{2}\pi\theta_*^2\right],
\label{eq:leakage}
\end{align}}where the first term represents the classical geometric stellar leakage. The second term can be associated to the instrumental leakage, which is related to the instrumental variable phase shift $\delta \phi(t)$. Both terms are multiplied by the parasitic factor $(\epsilon-\epsilon')^2$, which modifies the leakage level.\\
In this section, we showed that the effect of the crosstalk phenomenon is to modify the null output including the off-axis flux coming the astrophysical source. Two important consequences follow from that. First, the detection of the off-axis signal is likely to be impaired because of the decrease of the off-axis transmission of the interferometer. Second, even though the off-axis source is detectable in terms of signal-to-noise ratio, its genuine signal is anyway modified and the corresponding estimation will be corrupted here by the factor $(\epsilon-\epsilon')^2$.  
\section{Noise analysis} \label{s:snr}
In this section, we derive the impact of crosstalk on the detection performances of a nulling interferometer, using our theoretical formalism and then numerical simulations. 

\subsection{Theoretical determination}
To estimate the detection efficiency of a nulling interferometer affected by crosstalk, we thus define a signal-to-noise ratio taking into account the different noise contributions related to the stellar leakge and the background emission. We assume that the stellar leakage is perfectly calibrated by a rotation of the interferometer, to remove the geometric leakage, and by a calibrator observation, to evaluate the contribution of instrumental leakage. This procedure assumes an instrument stable enough between the calibrator and source observations, implying in particular a constant crosstalk level.
Therefore the modified null output (see Eq.~\ref{eq:finalnull}) is mostly affected by the photon noise associated to the modified stellar leakage (see Eq.~\ref{eq:leakage}), and the photon noise of the background emission. 
\begin{figure}
\centering
		\includegraphics[width=8cm]{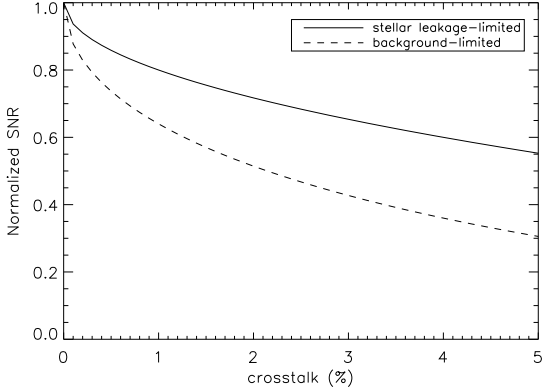}
		\caption{Evolution of the modified signal-to-noise ratio (SNR) of an off-axis source, as a function of the crosstalk level $\epsilon'^2$, in stellar leakage-limited and background-limited regimes. This signal-to-noise ratio is normalized to the SNR value without any flux contamination.}
		\label{fig:nullingSNR}		
\end{figure}
Our signal-to-noise ratio estimator is the ratio between the useful `signal', i.e. the flux of the off-axis source that should be measured at the destructive output during an integration time $\Delta t$, and the photon noise associated to the stellar leakage and the background emission, measured during the same integration time $\Delta t$ :
{\small \begin{align}
\nonumber
{\rm Signal}&=F_{\rm off}(\lambda)\Delta t \int\int{\tilde{R}_{\rm \lambda}(\theta,\alpha)f(\theta,\alpha) \theta d\theta d\alpha}, \\
\nonumber
{\rm Noise}&=\sqrt{\sigma_{B_{*}}^2(\lambda)+\sigma_{B_{bckg}}^2(\lambda)},
\end{align}}where $\sigma_{B_{*}}^2(\lambda)=F_*(\lambda)\Delta t (\epsilon-\epsilon')^2\left(\frac{\pi^3b^2\theta_*^4}{2\lambda^2}+\frac{<\delta\phi^2>}{2}\pi\theta_*^2\right)$ is the variance of the photon noise associated to the stellar leakage, $\sigma_{B_{bckg}}^2(\lambda)$ is the variance of the photon noise associated to the background emission measured during the integration time $\Delta t$. $F_{\rm off}(\lambda)=B_{\rm off}(\lambda)S_{\rm tel}T(\lambda)\Delta\lambda$ and $F_{\rm *}(\lambda)=B_{\rm *}(\lambda)S_{\rm tel}T(\lambda)\Delta\lambda$ are the total power of the off-axis source and the star, respectively, collected by an interferometer having telescopes of surface $S_{\rm tel}$, an overall instrumental transmission $T(\lambda)$, within a spectral bandwidth $\Delta\lambda$.  $<\delta\phi>^2$ is the quadratic temporal mean of the instrumental phase shift error during the same integration time $\Delta t$. Assuming that a fringe tracker is used, we can reasonably consider that the mean of $\delta\phi(t)$ is negligible against its standard deviation. Therefore, we get $<\delta\phi>^2\approx\sigma^2_{\delta\phi}$, with $\sigma_{\delta\phi}$ the standard deviation of the instrumental phase shift error $\delta \phi(t)$, during the integration time $\Delta t$.     
Then our estimator writes as :
{\footnotesize \begin{equation}
{\rm SNR}=\frac{F_{\rm off}(\lambda)\Delta t \int\int{\tilde{R}_{\rm \lambda}(\theta,\alpha)f(\theta,\alpha) \theta d\theta d\alpha}}{\sqrt{F_*(\lambda)\Delta t (\epsilon-\epsilon')^2\left(\frac{\pi^3b^2\theta_*^4}{2\lambda^2}+\frac{\sigma_{\delta\phi}^2}{2}\pi\theta_*^2\right)+\sigma_{B_{bckg}}^2(\lambda)}}
\label{eq:SNR}
\end{equation}}Using Eqs.\ref{eq:responseinstrumental} and \ref{eq:nophaseresponsepi}, we can write: $\tilde{R}_{\lambda}(\theta,\alpha)=2(\epsilon-\epsilon')^2R_{\lambda}(\theta,\alpha)$. Then, Eq.\ref{eq:SNR} becomes: 
\begin{itemize}
\item in stellar leakage-limited regime: 
{\small \begin{equation}
{\rm SNR}\approx(\epsilon-\epsilon')\frac{F_{\rm off}(\lambda)\Delta t \int\int{R_{\rm \lambda}(\theta,\alpha)f(\theta,\alpha) \theta d\theta d\alpha}}{\sqrt{F_*(\lambda)\Delta t\left(\frac{\pi^3b^2\theta_*^4}{2\lambda^2}+\frac{\sigma_{\delta\phi}^2}{2}\pi\theta_*^2\right)}}.
\label{eq:SNRepsilonstellar}
\end{equation}}
In this case, the modified SNR directly depends on the factor $\epsilon-\epsilon'$.\\
\item in background-limited regime: 
{\small \begin{equation}
{\rm SNR}\approx(\epsilon-\epsilon')^2\frac{F_{\rm off}(\lambda)\Delta t \int\int{R_{\rm \lambda}(\theta,\alpha)f(\theta,\alpha) \theta d\theta d\alpha}}{\sigma_{B_{bckg}}(\lambda)}.
\label{eq:SNRepsilonbackground}
\end{equation}}
In this case, the modified SNR directly depends on the factor $(\epsilon-\epsilon')^2$. 
\end{itemize}
Considering the coherent normalization of the beams, namely $\epsilon+\epsilon'=1$, Fig.~\ref{fig:nullingSNR} shows the evolution of the SNR of an off-axis source as a function of the crosstalk level $\epsilon'^2$, in the two extreme regimes shown above. The SNR is simply normalized by the intrinsic SNR value, i.e. without any flux contamination ($\epsilon'^2=0\%$), in order to illustrate the decrease against $\epsilon-\epsilon'=1-2\sqrt{\epsilon'^2}$ and $(\epsilon-\epsilon')^2=(1-2\sqrt{\epsilon'^2})^2$.  
For a crosstalk level of 1\%, the SNR is significantly reduced by 20\% and 36\%  in the stellar leakage-limited and background-limited regimes, respectively. This larger SNR decrease in the background-limited case illustrates 
the fact that the background emission is not affected by the intensity response of the interferometer (see Eq.\ref{eq:flux}). As a consequence, the background emission level is not modified in presence of crosstalk while the off-axis transmission is. In a background-dominated regime, the SNR is thus more sensitive to the level of crosstalk.  
\subsection{Numerical simulation}
\begin{figure}
\centering
		\includegraphics[width=9cm]{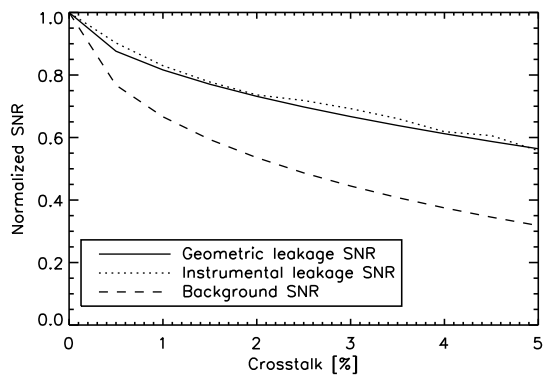} 
		\caption{Normalized SNR of a space-based Bracewell interferometer simulated with \emph{GENIEsim}, in stellar leakage-limited (geometric or instrumental leakage) and background-limited regimes. The SNR evolution is given with respect to the crosstalk level ($\epsilon'^2$). The SNR value is normalized by its value without crosstalk.}
		\label{fig:simu-SNR}		
\end{figure}

In order to apply our theoretical approach to a more realistic context, we used the \emph{GENIEsim} simulation software \citep{2006A&A...448..787A}, which was designed to simulate various Bracewell interferometer concepts such as the GENIE instrument at the VLTI. \emph{GENIEsim} has the advantage to have been extensively validated by cross-checking with performance estimates done by industrial partners during the GENIE phase A study. It performs end-to-end simulations of nulling interferometers, including the simulation of astronomical sources (star, circumstellar disc, planets, background emission), atmospheric turbulence (piston, longitudinal dispersion, wavefront errors, scintillation), as well as a realistic implementation of closed-loop compensation of phase and intensity perturbations by means of fringe tracking and wavefront correction systems. The output of the simulator basically consists of time series of photo-electrons recorded by the detector in the constructive and destructive outputs of the nulling combiner. The individual signal and noise contributions of the final output are extensively described in \citet{2006A&A...448..787A} and \citet{2008A&A...490..435D}. \\ 
In the context of this study, we have extended the use of \emph{GENIEsim} to handle crosstalk by following the mathematical description presented above. This was actually quite straightforward since only the theoretical expression of the complex amplitude of the transmission map (Eq.~\ref{eq:amplituderesponse}) had to be updated following Eq.~\ref{eq:electricfieldpar}. All output signal and noise contributions are then automatically taking crosstalk into account. Using \emph{GENIEsim}, we simulated a Bracewell interferometer with 1-m apertures separated by a 20-m baseline, and observing at 10~$\mu$m. We estimated, for different values of crosstalk, the broadband SNR that would be obtained by such an instrument when observing a hot Jupiter-like exoplanet orbiting, at 0.5~AU, a M5 star 5~pc away. For sake of comparison with our theoretical results,  we considered a normalized SNR and significant crosstalk level values up to 5\%. Fig.~\ref{fig:simu-SNR} shows the normalized broadband SNR obtained by \emph{GENIEsim} with respect to the crosstalk level ($\epsilon'^2$), in the case of three noise regimes, geometric leakage-limited, instrumental leakage-limited, and background-limited. In all three cases, the decrease of the normalized SNR provided by \emph{GENIEsim} follows in very good agreement our theoretical predictions (see Fig.~\ref{fig:nullingSNR}). For instance, a crosstalk level of 1\% leads to a relative SNR decrease of approximately 20\% in stellar leakage-limited regime (geometric or instrumental leakage-limited), and approximately 36\% in background-limited regime. This shows that our theoretical description can predict the impact of crosstalk even in more realistic conditions of phase perturbations, provided by \emph{GENIEsim}, than those considered in our equations. 

\subsection{Laboratory measurements}
A legitimate question is whether the level of crosstalk used in this study corresponds to realistic values expected for nulling interferometers. In the framework of nulling interferometry testbeds such as PERSEE \citep[see e.g.][]{2008SPIE.7013E..53J}, parasitic flux was actually used to feed metrology sensors. However, very low incoherent crosstalk levels of about $10^{-7}$ were estimated from a ZEMAX model \citep{Jacquinot2010}. The coherent crosstalk levels considered in our simulations (up to 5\%) thus overestimate what is currently measured in that case. The phenomenon of parasitic interference  is hence not an issue for such an experiment.  
However, it might be more problematic for long-term flagship missions dedicated to the detection of Earth-like planets for which a deeper null will be required. Therefore, a careful attention will anyway have to be paid to the possible sources of crosstalk, and especially coherent crosstalk, in the design of future nulling interferometry experiments. 

\section{Conclusion}\label{s:conclusion}

This work describes the phenomenon of parasitic interference in nulling interferometry as a consequence of coherent crosstalks. It can result from parasitic reflections inside the transmitting optics and/or from mutual beam contamination. Through an analytical approach, we have shown that this undesired effect affects the overall level of the transmission map of a nulling interferometer. In addition, it can possibly affect the observing and calibration procedures, which may not be yet fully defined for the future projects but would have to be considered. The two parameters involved in this degradation are the contamination rate - or crosstalk level - and the phase shift between each primary beam and its parasitic component.\\ 
An equal crosstalk between the beams result in a flattening of the transmission map at the destructive output, by a factor $(\epsilon-\epsilon')^2$, in the extreme case of a $\pi$ phase shift between the primary and secondary - or parasitic - beams. Here, $\epsilon^2$ represents the flux fraction of each primary beam, whereas $\epsilon'$ corresponds to the flux fraction of their corresponding parasitic beam, i.e. the crosstalk level or contamination rate. Considering in addition a differential crosstalk $\Delta\epsilon$ between the beams, the transmission map level is modified as $(\epsilon_1-\epsilon_1'+\Delta\epsilon)^2$, where  $\epsilon_1^2$ and $\epsilon_1'^2$ corresponds to the beam 1 in the `equal crosstalk' case. Because of this modification of the transmission map, the flux coming from the finite-size on-axis stellar source may be better rejected, while the flux of the off-axis source may be less transmitted. Therefore, in an equal crosstalk case, the photon noise due to stellar leakage is reduced by a factor $(\epsilon-\epsilon')$ while the off-axis astrophysical flux is reduced by a factor $(\epsilon-\epsilon')^2$. Since it is independent of the interferometer intensity response, the background emission level is not modified. This implies a degradation of the final null output SNR, by a factor $(\epsilon-\epsilon')$, in a stellar leakage - limited regime, and a greater degradation by a factor $(\epsilon-\epsilon')^2$, in a background-limited regime. To some extent, the estimation of the true astrophysical signal of the off-axis source, namely its spectrum, is therefore corrupted by the crosstalk effect, even though it is detectable. In addition, another important aspect is related to the retrieval of the astrophysical signal in actual observations requiring calibration procedures. For instance, since crosstalk impairs stellar leakage, the transfer function of a nulling interferometer will be biased if the calibration stars do not have the same angular diameter than the science star. Hence, to minimize the corruption of the true astrophysical signal, a careful attention should be paid to: 1) the design of the interferometer; in the context of the future VLTI instrument MATISSE \citep{2006SPIE.6268E..31L}, \citet{2009ApJ...706.1299M} proposed different solutions to prevent crosstalk and thus parasitic interference between beams, especially the separation of the path of each beam by a careful baffling inside the instrument; 2) the calibration procedures, involving the quality of calibration stars and the instrument stability especially in terms of crosstalk level.\\  
 We then compared our analytical study with numerical simulations of the impact of parasitic interference on the null output and the SNR delivered by a Bracewell interferometer. For that, we adapted the \emph{GENIEsim} simulation software to handle crosstalk. Our results show that the relative decrease of the SNR provided by \emph{GENIEsim} is in very good agreement with our mathematical description. A crosstalk of about 1\% implies a 20\% drop of the SNR in the geometric and instrumental leakage limited-regimes, while it implies a larger decrease of about 36\% of the SNR in the background-limited regime.\\ 
As previously mentioned, the parasitic interference phenomenon would appear to be negligible in current nulling interferometry testbeds such as PERSEE, where very low incoherent crosstalk levels of the order of $10^{-7}$ have been measured.\\
As a final conclusion, it appears that, up to now, little attention has been paid to the phenomenon of parasitic interference. This issue has been here formalized in the case of a Bracewell scheme. Notably, we could see that the detection of astrophysical objects providing weak signatures and flux level in the null output, such as hot Jupiter-like extrasolar planets, requires careful attention to various fine instrumental effects such as parasitic interference. Our work contributes to a better understanding of the factors optimizing the design of future planet-detecting interferometers like the NASA space nulling interferometer mission concept, FKSI \citep{2003ESASP.539...83D}. 
\section*{Acknowledgments}
We would like to thank the anonymous referee for his comments that helped to improve significantly the manuscript.
\bibliography{spieparasitic}   
\label{lastpage}
\end{document}